\documentclass[11 pt,a4paper,usenames,dvipsnames]{article}
\usepackage[utf8]{inputenc}
\usepackage{amsmath}
\usepackage{amsfonts}
\usepackage{amssymb}
\usepackage{pgfplots}
\usepackage{xcolor}

\usepgfplotslibrary{polar}
\usepackage{tikz}
\usepackage{subcaption}
\usepgfplotslibrary{fillbetween}
\usetikzlibrary{patterns}
\usepackage{indentfirst}
\usepackage{amsmath}

\author{Dr. Jacques Balayla MD, MPH, CIP, FRCSC\footnote{To whom correspondence should be addressed: Dr. Jacques Balayla MD, MPH, CIP, FRCSC. Quilligan Scholar. e-mail: jacques.balayla@mail.mcgill.ca. Osler Fellow. Department of Obstetrics and Gynecology. Faculty of Medicine. McGill University, Montreal, Quebec, Canada}}

\title{On the Formalism of The Screening Paradox}

\date{}

\begin{document}
\maketitle  
\begin{abstract}

Bayes' Theorem imposes inevitable limitations on the accuracy of screening tests by tying the test's predictive value to the disease prevalence. The aforementioned limitation is independent of the adequacy and make-up of the test and thus implies inherent Bayesian limitations to the screening process itself. As per the WHO's $Wilson-Jungner$ criteria, one of the prerequisite steps before undertaking screening is to ensure that a treatment for the condition screened exists. However, in so doing, a paradox, henceforth termed the ``screening paradox", ensues. If a disease process is screened for and subsequently treated, its prevalence would drop in the population, which as per Bayes' theorem, would make the tests’ predictive value drop in return. Put another way, a very powerful screening test would, by performing and succeeding at the very task it was developed to do, paradoxically reduce its ability to correctly identify individuals with the disease it screens for in the future. Where $J$ is Youden's statistic (sensitivity $a$ + specificity $b$ - 1), and $\phi$ is the prevalence, the ratio of positive predictive values at subsequent time $k$, $\rho(\phi_{k})$, over the original $\rho(\phi_{0})$ at $t_0$ is given by:

\begin{large}
\begin{equation}
\zeta(\phi_{0},k) = \frac{\rho(\phi_{k})}{\rho(\phi_{0})} =\frac{\phi_k(1-b)+J\phi_0\phi_k}{\phi_0(1-b)+J\phi_0\phi_k}
\end{equation}
\end{large} 

In this manuscript, we explore the mathematical model which formalizes said screening paradox and explore its implications for population level screening programs. In particular, we define the number of positive test iterations (PTI) needed to reverse the effects of the paradox as follows:
\begin{large}
\begin{equation}
n_{i\phi_e}=\left\lceil\frac{ln\left[\frac{\omega\phi_e\phi_k-\omega\phi_e}{\omega\phi_e\phi_k-\phi_k}\right]}{2ln\omega}\right\rceil
\end{equation}
\end{large} 
where $\omega$ is the square root of the positive likelihood ratio (LR+).
\end{abstract} 
\newpage

\section{Bayes' Theorem and Predictive Values}

Bayes’ Theorem describes the probability of an event occurring based on prior knowledge of conditions related to that specific event \cite{balayla2020prevalence}. The essence of the Bayesian approach is to provide a mathematical model explaining how existing beliefs change in light of new evidence. Remarkably, Bayes theorem' has applications in innumerable fields. Indeed, it has significant implications in epidemiological modelling as well. From Bayes' theorem, we can derive the positive predictive value $\rho(\phi)$ (PPV) of a screening test, defined as the percentage of patients with a positive screening test that do in fact have the disease, as follows\cite{balayla2020prevalence}:

\begin{large}
\begin{equation}
\rho(\phi) = \frac{a\phi}{ a\phi+(1-b)(1-\phi)} 
\end{equation}
\end{large} 
\\
\
where $\rho(\phi)$ = PPV, a = sensitivity, b = specificity and $\phi$ = prevalence.
\\
\

The PPV $\rho(\phi)$ is therefore a function of the disease prevalence, $\phi$. As the prevalence increases, $\rho(\phi)$ also increases and vice-versa \cite{brenner1997variation}.

\section{Bayesian Dynamics of Predictive Values}

Let us define a hypothetical disease present in a population. Said condition has a preclinical phase and is amenable to screening through a given test designed to detect it. The test therefore has all of the pertinent screening parameters - sensitivity, specificity, and negative and positive predictive values. Finally, as required by the WHO's $Wilson-Jungner$ criteria, a treatment for the condition screened exists. Let us denote $\phi_0$ as the original or initial prevalence before screening is undertaken. As per equation (3), we thus obtain a positive predictive value $\rho(\phi_0)$ at $t_0$:

\begin{large}
\begin{equation}
\rho(\phi_0) = \frac{a\phi_0}{ a\phi_0+(1-b)(1-\phi_0)} 
\end{equation}
\end{large} 

It follows that as the individual with a positive screening test is treated, the disease prevalence $\phi_0$ drops by some magnitude $k$, which represents the percentage reduction in prevalence. Consequently, the predictive value $\rho(\phi)$ will drop by some factor as well, so that individuals who test positive at some time $t_{x>0}$ experience a positive predictive value of:

\begin{large}
\begin{equation}
\rho(\phi_{0}-k) = \frac{a(\phi_{0}-k)}{ a(\phi_{0}-k)+(1-b)[1-(\phi_{0}-k)]} 
\end{equation}
\end{large} 
\

\newpage

From the above equations, we define the ratio $\zeta$ as follows:

\begin{large}
\begin{equation}
\rho(\phi_{0})>\rho(\phi_{0}-k) \Rightarrow \frac{\rho(\phi_{0}-k)}{\rho(\phi_{0})} = \zeta(\phi_{0},k) 
\end{equation}
\end{large} 

where $0<\zeta<1$ and $k>0$. Given the shape of the screening curve, and the principle of the prevalence threshold, even small changes in the prevalence $\phi$ can have significant changes in the positive predictive value  $\rho(\phi)$. To determine the degree of reduction in the predictive value of the screening test at time $t_k$, we take the ratio of $\rho(\phi)$ at two different times, be it $t_0$, and some later time $t_k$ with a prevalence reduction of $\phi_{0}-k$, where $k<\phi$ is the percentage reduction in prevalence:
\\
\
\begin{large}
\begin{equation}
 \frac{\rho(\phi_{0}-k)}{\rho(\phi_{0})}  = \frac{\left(\phi-k\right)\left[a\phi+\left(1-b\right)\left(1-\phi\right)\right]}{\phi\left[a\left(\phi-k\right)+\left(1-b\right)\left(1+k-\phi\right)\right]}
\end{equation}
\end{large} 
\

Since $\phi_{0}-k$ yields a new, lower prevalence $\phi_{k}$, we can re-write the above equation as:

\begin{large}
\begin{equation}
 \frac{\rho(\phi_{k})}{\rho(\phi_{0})}  = \frac{\phi_{k}\left[a\phi_0+\left(1-b\right)\left(1-\phi_0\right)\right]}{\phi_0\left[a\phi_{k}+\left(1-b\right)(1-\phi_{k})\right]}
\end{equation}
\end{large} 
\

Expanding the parentheses and simplifying the expression where the sum of the sensitivity $a$ and specificity $b$ is defined by $\varepsilon = a+b$, we obtain:

\begin{large}
\begin{equation}
\zeta(\phi_{0},k) = \frac{\rho(\phi_{k})}{\rho(\phi_{0})} =\frac{\phi_k(1-b)+\phi_0\phi_k (\varepsilon-1)}{\phi_0(1-b)+\phi_0\phi_k (\varepsilon-1)} 
\end{equation}
\end{large} 
\

The term $\varepsilon-1 = a+b-1$ has been previously defined in the context of receiver-operating characteristics (ROC) curves, and is termed the Youden's $J$ statistic \cite{youden1950index}. As such, we can re-write the above equation as:
\begin{large}
\begin{equation}
\zeta(\phi_{0},k) = \frac{\rho(\phi_{k})}{\rho(\phi_{0})} =\frac{\phi_k(1-b)+J\phi_0\phi_k}{\phi_0(1-b)+J\phi_0\phi_k} 
\end{equation}
\end{large} 
\
From the above relationship, we infer:
\begin{large}
\begin{equation}
\lim_{k \to 0}\frac{\rho(\phi_{k})}{\rho(\phi_{0})}=\lim_{k \to 0}\frac{\rho(\phi_{0}-k)}{\rho(\phi_{0})}=\lim_{k \to 0}\zeta = 1
\end{equation}
\end{large}

$\zeta(\phi_{0},k)$ may be considered as the predictive value percentage loss as the prevalence decreases from $\phi_0$ to $\phi_k$. If we consider both $\phi_0$ and $k$ as independent variables, both of which affect $\rho(\phi)$, we can establish the individual contributions of each variable towards $\rho(\phi)$ relative to each other through the partial differential equation $\zeta(\phi_0, k)$.

\newpage
\section{The Prevalence Threshold}
We have previously defined the prevalence threshold as the prevalence level in the screening curve below which  screening tests start to fail \cite{balayla2020prevalence}. In technical terms, this is equivalent to the inflection point in the screening curve below which the the rate of change of a test's positive predictive value drops at a differential pace relative to the prevalence\cite{balayla2020prevalence}. This value, termed $\phi_e$, is defined at the following point on the prevalence axis:
\

\begin{large}
\begin{equation}
\phi_e = \frac{\sqrt{a\left(-b+1\right)}+b-1}{(a+b-1)}=\frac{\sqrt{a\left(-b+1\right)}+b-1}{(\varepsilon-1)}
\end{equation}
\end{large}

The corresponding positive predictive value is given by plotting the above equation into the positive predictive value equation. In so doing we obtain:

\begin{large}
\begin{equation}
\rho(\phi_e)=\sqrt{\frac{a}{1-b}}\left[\frac{\sqrt{a\left(-b+1\right)}+b-1}{(\varepsilon-1)}\right]
\end{equation}
\end{large}

Interestingly, the above expression leads to the well known formulation for the positive predictive value as  a function of prevalence and the positive likelihood ratio (LR+), defined as the sensitivity $a$ over the compliment of the specificity $b$ \cite{streiner2019statistics}. 

\begin{large}
\begin{equation}
\rho(\phi_e)=\phi_e\sqrt{\frac{a}{1-b}}
\end{equation}
\end{large}

From the above relationship we can identify three different points in the screening curve, notably $\phi_0$, $\phi_k$, and $\phi_e$.
\\
\

\begin{center}

\begin{tikzpicture}
 
	\begin{axis}[
    axis lines = left,
    xlabel = $\phi$,
	ylabel = {$\rho(\phi)$},    
     ymin=0, ymax=1,
    legend pos = south east,
     ymajorgrids=false,
     xmajorgrids=false,
    grid style=dashed,
    width=6 cm,
    height=6cm,
     ]
	\addplot [
	domain= 0:1,
	color= blue,
	]
	{(0.85*x)/((0.85*x+(1-0.95)*(1-x))};
	\node[above,red] at (795,83) {$\phi_0$ = 0.38};
	\node[above,orange] at (795,73) {$\phi_e$ = 0.25};
	\node[above,ForestGreen] at (795,63) {$\phi_k$ = 0.18};
	\addlegendentry{$\varepsilon = 1.75$}
	\end{axis} 
	 \draw [red, thick] (1.65,0) -- (1.65,4.02);
	 \draw [orange, thick] (1.15,0) -- (1.15,3.78);
	  \draw [ForestGreen, thick] (0.80,0) -- (0.80,3.48);
\end{tikzpicture}
\end{center}
\begin{center}
\begin{small}
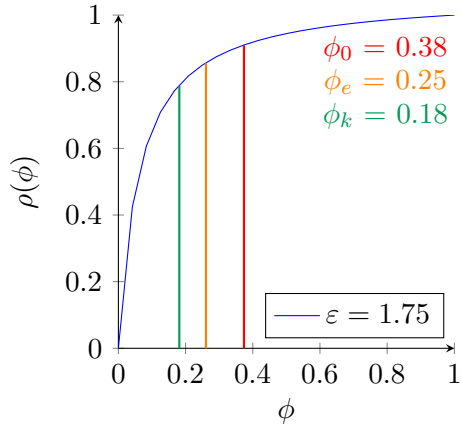
\textbf{Figure 1}. Example illustration of $\phi_0$, $\phi_k$, and $\phi_e$ where a = 0.85, and b = 0.90. 
\end{small}
\end{center}
\

We can deduce important relationships between these values that contextualize the screening paradox. First, we observe that though by definition $\phi_0>\phi_k$, $\phi_e$ can either be outside or in between $\phi_0$ and $\phi_k$. As such three different scenarios may arise. Herein we explore each.

\subsection{First scenario: $\phi_e>\phi_0>\phi_k$}

By design, $\phi_0>\phi_k$. Let $\phi_e$ define the prevalence threshold such that $\phi_e>\phi_0>\phi_k$. It then follows that $\phi_e-\phi_k>\phi_0-\phi_k$ and thus $\phi_e-\phi_0>0$. Since $\phi_0$ = $\phi_k+k$, we obtain $\phi_e-\phi_k-k>0$ or otherwise stated, $\phi_e>\phi_k+k$  and thus $\phi_e-\phi_k>k$. We thus infer that:
\begin{large}
\begin{equation}
\lim_{k \to \phi_0}\zeta(\phi) \sim 0
\end{equation}
\end{large}
and
\begin{large}
\begin{equation}
\lim_{k \to 0}\zeta(\phi) \sim 1
\end{equation}
\end{large}

\subsection{Second scenario: $\phi_0>\phi_k>\phi_e$}

This scenario is akin to the first scenario in that the prevalence threshold lies outside the range between $\phi_0$ and $\phi_k$. However, an important difference arises. While $\lim_{k \to 0}\zeta(\phi) \sim 1$, by design $\phi_k>\phi_e$, and thus the maximum value that $k$ can take cannot be greater than $\phi_0-\phi_e$, as $\phi_k\rightarrow\phi_e$. We thus infer that:
\begin{large}
\begin{equation}
\lim_{k \to \phi_0-\phi_e}\zeta(\phi) \sim 1
\end{equation}
\end{large}
and
\begin{large}
\begin{equation}
\lim_{k \to 0}\zeta(\phi) \sim 1
\end{equation}
\end{large}

The above relationships follow since for $\phi>\phi_e \rightarrow d\zeta/d\phi \sim 0$, as per equation (11).
\subsection{Third scenario: $\phi_0>\phi_e>\phi_k$}

Perhaps the most interesting scenario is one where $\phi_0>\phi_e>\phi_k$. By design $\phi_0>\phi_k$. Let $\phi_e$ define the prevalence threshold such that $\phi_0>\phi_e>\phi_k$. It then follows that $\phi_0-\phi_k>\phi_0-\phi_e$ and thus $0>\phi_k-\phi_e$. Since $\phi_k$ = $\phi_0-k$, we obtain $0>\phi_0-k-\phi_e$  and thus:

\begin{large}
\begin{equation}
k>\phi_0-\phi_e
\end{equation}
\end{large}
We thus infer that:
\begin{large}
\begin{equation}
\lim_{k \to \phi_0-\phi_e}\zeta(\phi) \sim 1
\end{equation}
\end{large}

In other words, when the initial prevalence lies beyond the prevalence threshold, changes in prevalence such that $k$ approaches the difference between $\phi_0-\phi_e$, the ratio of positive predictive values as determined by the $\zeta(\phi)$ function approaches 1.
However, we can theorize a case where $k$ is sufficiently large so that $\phi_k$ goes well below $\phi_e$ and therefore: 
\begin{large}
\begin{equation}
\lim_{k \to \phi_0}\zeta(\phi) \sim 0
\end{equation}
\end{large}
\section{The Screening Paradox at the Population-Level}
The mechanism by which the screening paradox arises is depicted through the following arrow flow diagram:
\\
\
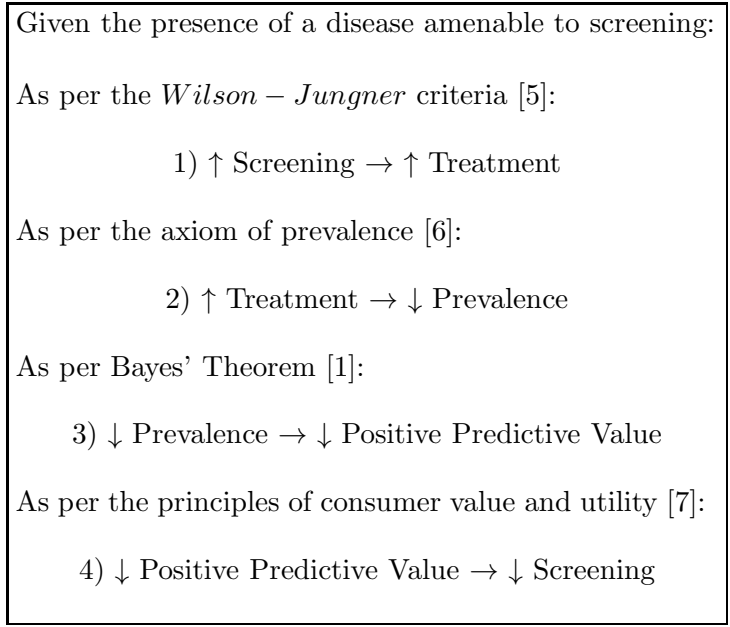
\begin{center}
\fbox{\begin{minipage}{24em}
Given the presence of a disease amenable to screening:
\\
\

As per the $Wilson-Jungner$ criteria \cite{wilson1968principles}:
\

\begin{center}
1) $\uparrow$ Screening $\rightarrow$ $\uparrow$ Treatment
\
\end{center}
As per the axiom of prevalence \cite{baldessarini1983predictive}:
\

\begin{center}
2) $\uparrow$ Treatment $\rightarrow$ $\downarrow$ Prevalence
\
\end{center}
As per Bayes' Theorem \cite{balayla2020prevalence}:
\

\begin{center}
3) $\downarrow$ Prevalence $\rightarrow$ $\downarrow$ Positive Predictive Value
\
\end{center}
As per the principles of consumer value and utility \cite{winsten1981competition}:
\

\begin{center}
4) $\downarrow$ Positive Predictive Value $\rightarrow$ $\downarrow$ Screening
\\
\

\end{center}
\end{minipage}}
\end{center}
\

Given the screening paradox, an increase in screening eventually leads to less, or more accurately lower quality, screening as the prevalence drops. This paradox is inherently insurmountable unless acted upon by a subsequent test - either the same test repeated serially or an altogether different, better test \cite{balayla2020bayesian}. 

\newpage
\section{Overcoming the Screening Paradox}
As the prevalence in a population drops with successful population-level screening and treatment, the positive predictive value of the screening test drops, and the false discovery rate, which is equivalent to the complement of the positive predictive value, increases. The aforementioned paradox occurs any time that disease is successfully treated because while $d\phi/d\rho$ drops throughout the function's domain, it never reaches 0. In other words, the positive predictive value function always increases throughout its domain, so even minute changes in prevalence will bring about changes in the positive predictive value. That said, as we described above, the critical factor is where lie the initial prevalence level $\phi_0$, the subsequent prevalence level $\phi_k$, their difference $k$, and how they relate to the prevalence threshold, $\phi_e$, below which the screening paradox becomes more pronounced. In the presence of a screening paradox it is worth considering potential solutions to overcoming the losses in predictive value as the prevalence drops. Though many options exist, the most logical step would be to undertake serial testing, be it with the same test undertaken serially or an alternative test altogether. Herein we explore both scenarios.

\subsection{Repeated testing with a single test}
We have shown in previous work that a screening test carried out serially improves the overall positive predictive value when each individual test iteration is positive \cite{balayla2020bayesian}. The number of serial iterations $n_i$ required to achieve a desired $\rho(\phi)$ is given by the following ceiling function:
\begin{large}
\begin{equation}
n_i =\lim_{\rho \to k}\left\lceil\frac{ln\left[\frac{\rho(\phi-1)}{\phi(\rho-1)}\right]}{ln\left[\frac{a}{1-b}\right]}\right\rceil 
\end{equation}
\end{large}
The key question then becomes, how many serial positive tests are needed to mitigate or reverse the effect of the screening paradox when $\phi_k<\phi_e$? In other words, to achieve a positive predictive value comparable to that under $\phi_0$? Given the geometry of the screening curve, the answer should be that the PPV ought to at least attain the level at the prevalence threshold as described in the third scenario in section 3.3 of this manuscript so that $\zeta(\phi)\sim 1$. 
We can calculate the number of iterations $n_{i\phi_e}$ needed using the formula above, by plugging $\rho(\phi_e)$ into $\rho$ as defined in equation (13),  where $\rho(\phi_e)=\phi_e\sqrt{\frac{a}{1-b}}$.

\begin{large}
\begin{equation}
n_{i\phi_e} =\left\lceil\frac{ln\left[\frac{\phi_e\sqrt{\frac{a}{1-b}}(\phi-1)}{\phi(\phi_e\sqrt{\frac{a}{1-b}}-1)}\right]}{ln\left[\frac{a}{1-b}\right]}\right\rceil 
\end{equation}
\end{large}

The above expression can be simplified by considering  the square root of the positive likelihood ratio $\sqrt{\frac{a}{1-b}}$ as $\omega$, and $\phi_k$ is the prevalence at subsequent time $k$ such that:

\begin{large}
\begin{equation}
n_{i\phi_e} = \left\lceil\frac{ln\left[\frac{\omega\phi_e(\phi-1)}{\phi(\omega\phi_e-1)}\right]}{2ln\omega}\right\rceil = \left\lceil\frac{ln\left[\frac{\omega\phi_e\phi_k-\omega\phi_e}{\omega\phi_e\phi_k-\phi_k}\right]}{2ln\omega}\right\rceil
\end{equation}
\end{large}

We take the ceiling function of the above equation to ensure that we obtain an integer number of of positive test iterations (PTI) needed to surpass the prevalence threshold \cite{balayla2020bayesian}.

\subsection{Using a different screening test}
We can likewise revert the effects of the screening paradox by using a different screening test all together, which is the most common scenario in clinical practice today. That said, it would be impractical to determine the number of iterations of a different test for numerous reasons. First, different tests would have different sensitivity/specificity parameters, so a third test may be then needed in the rare scenario where two different positive ones are insufficient - rendering the notion of iteration inadequate. Likewise, and perhaps more importantly, there may not be an alternative screening test for a particular condition altogether, so the above exercise may be moot. 

\section{Conclusion}
In this manuscript, we explore the mathematical model which formalizes the screening paradox and explore its implications for population level screening programs in the three possible scenarios - each as a function of the position of the initial prevalence of a condition  relative to the prevalence threshold level of its screening test. Likewise, we provide a mathematical model to determine the predictive value percentage loss as the prevalence decreases and define the number of positive test iterations (PTI) needed to reverse the effects of the paradox when a single test is undertaken serially. Given their theoretical nature, clinical application of the concepts herein reported need validation prior to implementation.  
\newpage

\bibliographystyle{unsrt}
\bibliography{references}
\end{document}